# The development and implementation of a PhD Program in ICT for the Kosovo Education System

Anita MIRIJAMDOTTER[1], Krenare PIREVA NUCI[2], Michele GIBNEY[3,1], Patrik ELM[1]
[1]Linnaeus University, Faculty of Technology
[2]University for Business and Technology, Faculty of Computer Science and Engineering
[3]University of the Pacific
anita.mirijamdotter@lnu.edu, krenare.pireva@ubt-uni.net, mgibney@pacific.edu,
patrik.elm@lnu.edu

**Abstract**

Despite ever accelerating workplace changes, including rapidly expanding technological access and fast improving information and communication systems, the education system in Kosovo is not fully developed enough to provide a high-quality research-based education in Information and Communication Technology. Coping simultaneously with varied national priorities, Kosovo – a small country with 2 million inhabitants and a national budget of only 2.3 billion – lacks the needed investments to fundamentally transform the quality of the education system. A funded ICT doctoral program would address today's workforce priorities and requirements. The design and delivery of a national PhD program in ICT is crucial for Kosovo in order to ensure competitive readiness within the regional education systems and national economies of the West Balkans - and beyond. This paper argues the need for PhD programs and offers insights into a proposed project, the aim of which is to put Kosovo on the map by offering a PhD in the ICT field.

## 1. Introduction

According to the Kosovo Agency of Statistics in 2018 [1], the average age in Kosovo is 30 whereas the unemployment rate for youth (15-24 years), that is, young people who are neither employed, nor in school or training, is 52.7%. In the same year Forbes [2] reported that ICT sector expertise, ranging from data scientists to information security officers, is in high demand and there are additional reports which state that demand will continue to increase exponentially [3]. The Kosovo IT strategy states that the strong average of young people makes Kosovo a runner-up in outsourcing working power to Eastern Europe. Additionally, having a high quality education in the ICT domain, will translate to Kosovo's ability to export those with high expertise in IT services, providing a long-term benefit for the economic growth of the nation and the region. The knowledge-based economy will speed up through technology transfer and creation of jobs with higher incomes [4].

To increase the academic level and quality of the education system in this transition nation - which celebrated its 10th anniversary in 2018 - continual investments are needed. Coping simultaneously with a variety of national priorities, Kosovo, as a small country with around 2 million inhabitants and a poor national budget of around 2.3 billion, has experienced constraints on the needed investments to drastically transform the quality of the education system. Further, research indicates that a continual investment in education brings dynamic potential for growth and innovation in emerging economies [5]; it is at such a lynchpin moment that Kosovo is currently mired and, thus, has a pressing need but lack of resources.

In this situation, the system of university education in Kosovo is experiencing an obstacle in terms of lack of focused research collaboration and output, which may be due to the lack of programs, mainly those at the PhD level in ICT [6]. The absence of such a graduate program in this transition nation makes it nearly impossible to create the national human intellectual capacity and technological infrastructure required to become a nation that fully integrates the ideas of the digital age and a knowledge society. This situation is usually experienced in small countries (less than 5 million inhabitants), where the lack of financial and human resources limits the ability to develop and implement high quality programs [7].

It is the aim of this project to implement a new, national curricula in ICT which will be realized by Kosovo local universities so they can profile them accordingly. Initially three higher education institutions (HEI) have been selected for the initial phase, and the aim is to increase the profiled number of other HEI in the future. The initial three are the University of Prishtina (UP), University for Business and Technology (UBT), and Ukshin Hoti University (UHU). Ministry of Education, Science and Technology in Kosovo (MESTK), will support the project implementation on the national level, as well as the Accreditation Agency of Kosovo (AAK). The latter will provide the requirements that need to be matched with respect to local legislation during the accreditation process. Also, the link between industry and academia will be established through the involvement of local industry as associated partners.

This paper report on the research conducted leading to the identification of the need for heightened competence for computer sciences students in Kosovo, then presents the project consortium including local university partners, and, finally, reports on the steps envisioned to establish a Kosovo national PhD program in ICT.

## 2. Background Information

Despite ever accelerating workplace changes, including rapidly expanding technology access and tools and fast improving information and communication systems, the education system in Kosovo is not yet fully developed to provide the high quality research education necessary to produce a qualified and competent ICT workforce.

The latest research that have been conducted from the Kosovo Association of Information and Communication Technology reported that the demand for heightened competence for computer sciences students in Kosovo and in the region is very high and those who finish their studies successfully can find work in the labour market. However, employers report that these students lack the soft skills and problem solving abilities essential for workplace success in the industrial sector [8].

This situation drove the project team to further research and investigate what was causing the discrepancy. Current state factors that lead to excerbating this situation were analyzed and three discrepancies were identified: People, Programs and Infrastructure.

- *People*: This includes, for example, educators and ICT workers in the field who demonstrate a lack of adequate competencies in the subject matter. Partially this could be due to a lack of high quality research production - particularly in HEI, due to the missed opportunity of PhD students conducting research.
- *Programs*: None of the universities within Kosovo provide doctoral level schools in ICT. At one time, University of Prishtina did have a PhD program in Computer Engineering, however the program was unable to succeed due to a lack of full professors in this specific field.
- *Infrastructure*: As a transition nation, Kosovo infrastructure has been built gradually for the past 10 years but is not at the level to adequately support doctoral studies with elements such as research centers and institutes, field-specific equipment, and trained experts to demonstrate the equipment and support its use.

These three are the identified gaps which can be addressed by the creation of a national PhD program in ICT within Kosovo. This program will provide quality education at the PhD level, modeled by the international consortium in collaboration with the Kosovo partner institutions. An ICT research education program should address today's workforce priorities and requirements. These digital age knowledge, skills, and abilities reflect a global paradigm shift in professional qualifications and technology readiness, which serve to improve and enhance productivity and effectiveness. The initiative covers the three national priorities of Kosovo - quality in the education system, internationalization of programs, and increasing research capacities in the nation [9]. The design and delivery of a national PhD program in ICT is crucial for Kosovo in order to ensure competitive readiness within the regional education systems and national economies of the West Balkans - and beyond.

## 3. Consortium Details

The international consortium (Linnaeus University, Sweden; Norwegian University of Science and Technology, Norway; and South East European Center, Greece) support the initiative to develop and establish a common national PhD Program for ICT in Kosovo.

*Linnaeus University* (LNU) is one of the largest universities in Sweden with about 34,000 students and more than 2,000 staff members. Every academic year LNU hosts about 1,500 international students and staff members from more than 700 partner universities. LNU has a long experience in international programs for staff and students. Research at LNU is of high quality, nationally as well as internationally, and covers a wide range of disciplines. The research education at LNU that relates to this project is organized in the subject of Computer and Information Sciences, which spans three subjects: Computer Science, Media Technology, and Informatics. In a recent evaluation, made in 2018 by the Swedish Council for Higher Education, the conclusion was that LNU research education in Computer and Information Sciences is of high quality [10].

Linnaeus University is project coordinator and will be responsible for the overall coordination of project activities, including financial management, project monitoring, as well as corrective actions and continual reporting. All other local management, including project deliverables and related activities, are distributed to the consortium members.

*Norwegian University of Science and Technology* (NTNU) specializes in science and technology. With 40,200 registered students, 7,100 teaching and research staff, 3 campuses and offices in Brussels and Japan, NTNU is the largest university in Norway today with a history that dates back to 1910. NTNU's social mission is to create knowledge for a better world and deliver solutions that can change and improve everyday life. NTNU is Norway's most international university in terms of student exchange, both outgoing and incoming Erasmus students, and collaborates with more than 730 institutions in over 100 countries. A total of about 2,000 candidates are at any time enrolled in a PhD program at NTNU and there are about 250 PhD defenses every year and more than 40% of the candidates are non-Norwegian citizens. A very high share of NTNU's research is performed in cooperation with industrial partners or public and private enterprises. Key persons from NTNU to this project are members of boards of PhD programs who are responsible for developing the structure, content and quality of the doctoral education and are highly experienced lecturers and supervisors who regularly develop and teach courses [11]. NTNU is a partner in this project, and will transfer their extensive knowledge and experience in joint-research, joint supervision of Doctoral thesis, and will support the modelling and creation of the non-existing national PhD program that will serve for Kosovar universities.

*South East European Research Center* (SEERC) is an overseas research centre of the University of Sheffield, established as a non-profit legal entity in Thessaloniki, Greece. The Center was founded by City College, the University of Sheffields's International Faculty, in 2004. It is conducting multidisciplinary research in the fields of Enterprise, Innovation & Development, Information & Communication Technologies, and Society & Human Development. Being part of the University of Sheffield provides high quality experience and skills in innovative learning and teaching methods with strong expertise in Business and Economics, Computer Science, Information Technologies, Innovation Management, Psychology and a strong know-how regarding effective curriculum design. SEERC, since 2004, has implemented the "SEERC Doctoral Program" of the University of Sheffield which aims to bring high quality academic doctoral training in various disciplines (including ICT) to South East Europe. The SEERC Doctoral Program involves rigorous doctoral training and a supervision scheme that allows local and Sheffield based faculty to co-supervise PhD students [12].

SEERC will contribute its experience of developing a PhD program in cooperation with the staff of its parent institution, the University of Sheffield. This experience will help develop knowledge, procedures and capacity for the Kosovo ICT PhD program at all levels: setting research agendas, recruiting, developing skills, teaching, supervising and examining PhD students. SEERC's staff will also be involved in mentoring and training to increase research capabilities, and in the co-supervision of the first cohorts of the Kosovo PhD candidates, as part of the implementation of the project.

These international partners will contribute through co-design and transferring and contextualizing the 'know-how' expertise of PhD models to curricula customized to the unique situation and circumstances of this new sovereign nation. In this way, the Kosovo higher education system will design and implement a high quality national program delivered within a higher education infrastructure for the new generation of university students.

A high impact graduate program necessarily requires a robust academic and research culture, fortified by strong and enduring commitments to internationalization, innovation, and creativity. Therefore, the local experts (Kosovars who can situate learning) and the foreign expert partner institutions will also further joint supervision, joint research and related trainings organized in the project. The aim throughout will be to co-create a boundary crossing ecosystem that fosters and fortifies universities, faculty, staff and student inquiry.

Further, while this project will attempt to alleviate the absence of a PhD program in ICT, a secondary missing element in Kosovo education includes a Masters or PhD in Library and Information Science. Library staff from the international partners will be able to provide training to the doctoral candidates and the PhD program faculty in information literacy and information experience design to further the educational goals within the ICT curriculum. They will also be able to provide direction to vetted resources and advice to library staff at the Partner institutions.

Whereas, in Kosovo currently, there are 9 public and 22 private HEI in Kosovo [13]. Out of these 31 HEI, the University of Prishtina (UP) did previously offer a PhD program in the ICT field, however, due to a lack of human resources, this program is currently on hold; the re-accreditation process will be organised in the future. This is the only university in Kosovo that has ever offered PhD education in the ICT field.

For this project the local universities that are involved are:

*University of Prishtina* [14], 13 different faculties with 65 Programmes for Bachelor studies, 64 Programmes for Master studies and 17 Programmes for PhD studies. This year UP has reached 42,006 enrolled students, with 900 academic personnel and 300 administrative staff. Besides these facts, UP has a stable and updated curricula in Computer Engineering at both bachelor and master levels.

*University for Business and Technology* [15], 20 Faculties, with 43 study programs with 98 accredited Majors, 3 MBAs, 10 Joint International Study Programs, and 12 master programs (around 17,000 enrolled students in total). UBT has a stable and updated curricula in Computer Science, at both bachelor and master levels. These programs attract approximately 1,000 new students each year.

*Ukshin Hoti University* [16], currently more than 15,000 enrolled students. In total UHU offers 24 bachelor programs and 3 master programs.

## 4. Target Groups and Objectives

The ICT sector in Kosovo has created around 6,500 jobs in the past decade and, according to economic projections, another 10,000 new jobs will be created by 2025 [17]. This project aims to create long term effects, by generating high quality cadres for the education system in Kosovo, which will disseminate their knowledge to the young generations of ICT students and prepare them for the forthcoming growth of the ICT industry.

The target groups, and the needs to be addressed, include the following:

1. *Potential PhD Students* - targeted to receive a quality education that prepares them to work in the ICT field either in industry or academia;
2. *Faculty in higher education* - targeted to gain the appropriate level of education and support to pass on to the doctoral candidates;
3. *Staff in higher education* - targeted to understand their role in supporting the new national PhD program in ICT to the best of their ability;
4. *Administration in higher education* - targeted to understand their role in supporting the new national PhD program in ICT to the best of their ability;
5. *The participating higher education institutions* - targeted to attract and retain quality students, faculty, staff and administration to support the emerging national PhD program;
6. *Industry employers* - targeted to employ quality candidates in their workforce that have achieved the appropriate learning, skills and training in an accredited PhD program within Kosovo;
7. *Industry experts* - targeted to share their learning, skills, and abilities with the next generation in order to retain and bolster Kosovar knowledge in the ICT field.

The project aims to generate human capacities that will be able to adapt to dynamic working environment, either in industry or in higher education institutions in Kosovo. More precisely, the objectives of the project are:

1. Establish a national research school in ICT in Kosovo;
2. Establish a national curricula for PhD studies in the ICT field with key elements of learning outcomes, goals and skills;
3. Increase the research competences of Kosovo universities;
4. Facilitate the accreditation process of the individual universities' PhD programs to appropriate national bodies;
5. Implement the national PhD program and register the first generation of graduate students, which will be jointly supervised from Kosovo institutions and international institutions.

The project aims are inline with the Kosovo education strategy, formulated by the Ministry of Education, Science and Technology [18], which emphasizes the need for internationalization programs and increased research capacities. This project also addresses national priorities of Kosovo by adhering to the governance of strategic planning of HEI [9] for the purpose of meeting future needs of Kosovo's economic and societal development. In addition, it will target the prioritized quality assurance processes [4], which will be transferred to continuous curriculum feedback evaluation, by collaborating with LNU, NTNU and SEERC. Finally, the gist of the proposed program is aimed at developing research and innovation capacities, yet another national priority. The national and regional priorities of special attention are: internationalization of HEI, increase the quality of the education system, and increase the research capacities, which includes transfer of cultural values such as equity, democracy, and equal access to higher education [18].

## 5. Implementation Model

At the first stage, the project will analyze and research PhD programs in ICT globally. It will create activities through professional development workshops and training to increase the research qualifications of local staff professors. Finally, it will facilitate the process of establishing the national PhD program individually for the involved Kosovo university institutions, including all the required documentation and human capacities stated as prerequisites from the Accreditation Agency of Kosovo.

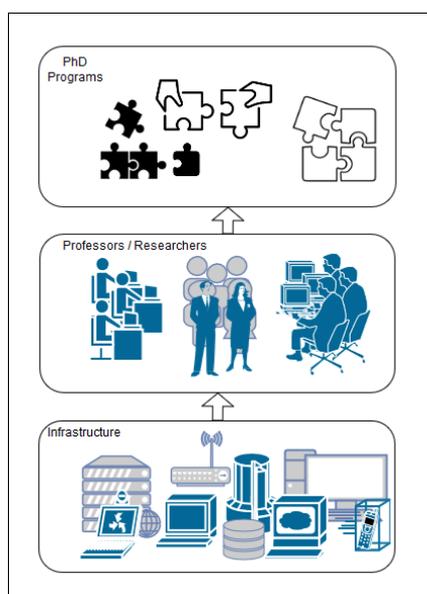

**Figure 1.** The proposal architecture for the PhD program

As shown in Figure 1, in order to create a national PhD program that could facilitate the local universities to be affiliated and offer their PhD programs individually, we have proposed a three layered methodology:

- The bottom layer, namely *the infrastructure layer*, is initially planned to use the virtual research infrastructure of Sweden, and in parallel to that, establish the local appropriate infrastructure to the selected local partner institutions.
- The middle layer, namely *the research capacity layer*, increases the local partner institution capacities through workshops, training and, beside this, through joint supervision and joint research projects. The human capacities and research competences are increased by mentorship and by the transfer of the international expertise. Additionally, by inviting the Kosovars that have finished their PhD studies abroad, in a number of seminars [19] local capacities of the partner institutions are further increased. Also, the first generation of the PhD students will be mentored from at least one international consortium institution Professor and in parallel to that, be co-mentored by a local partner institution.
- The upper layer, namely *the national PhD program*, will be modeled during this project with learning outcomes and appropriate courses, and implemented by local partner institutions as a framework through which they may develop their individual PhD programs. These latter should follow the prerequisites stated by the actual Kosovo regulations in order to be recognized by the Accreditation Agency [20, 21].

The three year PhD curricula in ICT (180 ECTS) will be modeled based on international, national, and industry standards, aligned with ACM Curriculum Recommandation [22]. The aim is to generate a well-designed, consistent, and coherent learning paths for first year courses in graduate Computer Science curricula, which integrate the suggestions of IEEE/ACM as well as other stakeholders'. During the development of the PhD program, the first year courses will be designed to offer problem oriented project activities, identified from business and industry, which will facilitate the learning process. For the last two years, the PhD program continues with the individual student's specific research project.

Conceptualizing the whole project as three layered architecture (see Figure 1), the aim is to create a sustainable national PhD program, where initially reliance exists on the international consortium institutions' - both physical and virtual infrastructure - and their expertise. Envisioning discussions see that the three year project scope will experience the establishment of appropriate local infrastructure and research capacities. All the activities are organized within 8 work packages, ranging from preparation, initiation, developing human capacity, developing the PhD program and the accreditation process. Further packages include quality assurance, dissemination and exploitation, and overall project management. The biggest efforts will be given in the development work.

In the longer term, along the project development, the project partners also plan on reaching the target groups at the completion of the project. The project establishes the PhD program and to disseminate the results of the project, the target groups, internal and external to the initial participating HEI, will be informed through advertising, social media, broadcasting, etc. In this respect, new local and regional partners will be crucial to help higher education institutions reach the target groups, and enable permanent program improvements.

In addition, policymakers, education authorities, the computer science industry, the research and teaching community team, but also students, will have a clear insight into the creation, implementation, and additional development of the national PhD program in Kosovo.

## 5. Conclusions

In Kosovo none of the public or private higher educational institutions have a PhD program in ICT. The idea of proposing a national PhD program in this domain, reliant on the expertise of universities that have demonstrated long time experience in similar programs, is an innovative idea that will establish an entrepreneurial program for the nation and for the region. The proposal to increase the research competences by joint supervision, mentorship and joint research projects among international partner institutions and Kosovo, and by initiating a national research education institute for, at this stage, University of Prishtina (UP), University for Business and Technology (UBT), and Ukshin Hoti University (UHU) and aiming to extend for other Kosovo HEI in the coming years, is the second innovative aspect.

Expected effects of this project are:
- Increased quality of research education in the ICT field in Kosovo
- Increased national quality assurance policy for research education in ICT in Kosovo
- Increased mobility of PhD students and researchers of Kosovo with partner institution counterparts
- Increased human capacities for research education in the ICT field in Kosovo

We foresee that the creation of new links with industry, the transfer of joint supervision and joint research know-how from international professors, and the virtual research education institute that will be facilitated by the partner countries' institutions, parallel to creation of the national PhD program, is an innovative and sustainable proposal, which will necessitate more attention in Kosovo universities. It is also in line with the emphasis of the Kosovo national priorities. Finally, we find the joint synergy of local universities, organized as parts of the virtual research education system for the benefit of accomplishing the national priorities for the public good, is an innovative character in Balkan countries.

These innovative attributes as part of the national PhD program in ICT will offer an education to newly enrolled students to be more innovative and creative thinkers in their particular sub domains, and simultaneously rigorous in their approach to problem solving.

At the end of the project the concrete results for Kosovo are as follows:
- The establishment of the national PhD research school
- The creation of appropriate documentation for the operation of the national PhD school (strategy, regulation, student handbook)
- The establishment of the PhD curricula in partner institutions.
- The successful affiliation of the initial partner institutions, and aiming to increase the number of affiliated partner institutions in the future.

Since this PhD program will be developed and implemented in Kosovo, the need for collaboration with experienced experts in international institutions is crucial. Here the international partners, such as LNU, NTNU, and SEERC, will jointly bring their expertise, contributing to European level added value, through various creative and innovative activities. An additional element for European added value, which will be of use during this project, is the assistance and cooperation of trained library and information professionals.

Since this EU initiative aims to further inclusion of western Balkans countries in the economic, social, and political life of Europe, UP, UBT and UHU depend on providers of established ICT programs to further advance the research study quality overall in Kosovo. Therefore, this initiative supports Kosovar graduates to gain the knowledge, skills, and abilities, which enable them to contribute to the workplace and the society at local, regional, and global levels.

**Acknowledgements**: We are deeply grateful to: Arianit Kurti, Welf Löwe, Marcelo Milrad, Linnaeus University; Marian Dema, Lule Ahmeti, University of Prishtina; Edmond Hajrizi, Edmond Jajaga, UBT; Nikos Zaharis, Petros Kefalas, SEERC; Ali Shariq Imran, Sule Yildirim Yayilgan, NTNU University; Rame Vataj, Ukshin Hoti University; and Mary Somerville, University of the Pacific, for their insightful comments and contributions to the project proposal on which this paper builds. Further we are grateful also to all Industry companies that supported this idea, as well as Ministry of Education, Science and Technology in Kosova, and Accreditation Agency in Kosova.

**References**

[1] Kosovo Agency of Statistics. (2018). Statistical Yearbook of the Republic of Kosovo 2018 [Online]. Available: http://ask.rks-gov.net/media/4369/statistical-yearbook-2018.pdf
[2] Forbes. (2018, November 27). Global Finance Leader Study: Data Science Is Most Important Expertise For Future Finance Team. *Workday BrandVoice* [Online]. Available: https://www.forbes.com/sites/workday/2018/11/27/global-finance-leader-study-data-science-is-most-important-expertise-for-future-finance-team/#3cc784a95575
[3] UBT Higher Education Institution.(2019). Available: https://www.ubt-uni.net
[4] IT Strategy Working Group. *Kosovo IT Strategy,* Republic of Kosovo, Government, Ministry of Economic Development [Online]. Available: https://stikk.org/wp-content/uploads/2018/11/Kosovo_IT_Strategy_V01-00_29-06-2016.pdf
[5] D. Lim. *Quality Assurance in Higher Education: A Study of Developing Countries*. Routledge (2008).
[6] List of Kosovo HEI accredited from Accreditation Agency. 2019. [Online]. Available: http://akreditimi-ks.org/docs/Downloads/Accreditation/kshc61/Accredited_01222019.pdf
[7] S. Z. Salas-Pilco and N. W. Law, "ICT Curriculum Planning and Development: Policy and Implementation Lessons from Small Developing States". In *ICT-Supported Innovations in Small Countries and Developing Regions*, 2018, pp. 77-98. Springer, Cham.
[8] STIKK, Kosovo Association of Information and Communication Technology. (2015). Mapping of ICT Sector Labour Supply and Demand [Online]. Available: https://stikk.org/wp-content/uploads/2018/11/Publications_2015_-_Skills_Gap_EN.pdf

[9]   Kosovo National Development Strategy 2016 - 2021. 2016. Available: http://www.kryeministri-ks.net/repository/docs/National_Development_Strategy_2016-2021_ENG.pdf
[10]  Linnaeus University (LNU), Sweden. Homepage Available: https://lnu.se/
[11]  Norwegian University of Science and Technology (NTNU), Norway. Homepage. Available: https://www.ntnu.edu/
[12]  South East European Research Center (SEERC), Greece. Homepage. Available: https://www.seerc.org/new/
[13]  List of Public and Private HEI in Kosovo. (2019). Available: http://www.akreditimi-ks.org/new/index.php/en/download/higher-eduacion-institutions
[14]  University of Pristina (UP), Kosova. Homepage Available: https://uni-pr.edu
[15]  UBT Higher Education Institution (UBT), Kosova. Homepage Available: https://ubt-uni.net
[16]  Ukshin Hoti University (UHU), Kosova. Homepage Available: https://uni-prizren.com/
[17]  A. Diakonidze, S. Raja, and N. Gelvanovska. (2016). Kosovo Digital Economy: Skills for Jobs [Online]. .Available: http://www.mzhe-ks.net/repository/docs/Kosovo_Digital_Economy-Digital_Skills_for_Jobs.pdf
[18]  Ministry of Education, Science and Technology. (July 2016). Kosovo Education Strategic Plan 2017-2021. Available: https://masht.rks-gov.net/uploads/2017/02/20161006-kesp-2017-2021-1.pdf
[19]  UNDP. (2018). Research Study into Brain Gain:Reversing Brain Drain with the Albanian Scientific Diaspora [Online]. Available: https://www.undp.org/content/dam/albania/docs/Brain_Gain%20web.pdf
[20]  Administrative Instruction (MEST). (2017). No 9/2017 for accreditation of Higher Education Institution in the Republic of Kosova [Online]. Available: https://masht.rks-gov.net/uploads/2017/08/pdfjoiner_1.pdf
[21]  Standard of Program Accreditation in Kosovo. (2019). Available: http://www.akreditimi-ks.org/new/index.php/en/accreditation/standards-of-accreditation
[22]  Association for Computing Machinery. (2019). Available: https://www.acm.org/education/curricula-recommendations

**Anita MIRIJAMDOTTER** is Professor, head of research, and subject responsible of Informatics at Linnaeus university since Jan. 2008. She is a board member of the Swedish National Research School, Management and Information Technology. Anita has supervised nine PhD students to completion, of which six as main supervisor, and eleven research students to Licentiate degree. She has served as opponent at eight PhD dissertations, including two internationals, and as member of twenty-nine PhD exam committees, including four internationals. Anita's research and teaching focus on design and management of information, communication, and decision systems in dynamic organizational settings. Her perspective includes: Systems Thinking Methodologies and Models, and Participatory Action Research. Anita has served as assessor for quality assurance of national and international educational programs, assessor of pilot projects in evaluation of Swedish research education, and assessor of education programs in own and related subjects. She has extensive experience delivering higher education across the globe, including countries in the Balkan Region, such as Kosovo and Albania. Additionally, she recently has been granted Erasmus + International Credit Mobility funds 2018-2020 for exchange of faculty, staff and students with University for Business and Technology, Kosovo, and University of Tirana, Albania. Please find more information on https://lnu.se/en/staff/anita.mirijamdotter/

**Krenare PIREVA NUCI** is full-time senior lecturer, in Faculty of Computer Science and Engineering at UBT (Kosova). Currently she is visiting lecturer in University of Tirana (Albania), in Applied Informatics, and mentoring Master Thesis in Linnaeus University (Sweden), in Faculty of Technology. She holds a PhD and Master degree in Computer Science, from University of Sheffield. And her main interests relay in Artificial Intelligence domain, and especially the applied techniques of AI in education domain. She has participated and published in a number of research paper in collaboration with number of Universities nationally and globally. Lately in UBT she is certified in IPMA International Project Management Association, Level D. Dr Pireva Nuci is active in organizing a number of national and international conference in the Computer Science domain.

**Michele GIBNEY** is a doctoral student in Informatics at Linnaeus University, Sweden, and currently working as the Head of Publishing and Scholarship Support at University of the Pacific in the United States. in 2019, she was awarded a United States State Department Fulbright Specialist one month residency to travel to Kosovo and support University for Business and Technology (UBT) in starting an institutional repository to collect, preserve and disseminate faculty, student and staff scholarly and creative works. The repository, the UBT Knowledge Center, https://knowledgecenter.ubt-uni.net/, is now live and contains over 1,500 records. She is also the recipient of an Erasmus+ grant for study abroad in Kosovo for three months. Michele's research focuses on the scholarly communication life-cycle - part of which is using ICT to access and disseminate scholarship globally. Please find more information on: https://works.bepress.com/michele_gibney/

**Patrik ELM** is assistant professor at the Linnaeus University, received his PhD in 2007 at Blekinge Institute of Technology, Computer Science. His skills and expertise are related to information quality in the context of the important role of information flows in organizations, in particular from the technical perspective and including information security concerns. Dr. Elms research and teaching is founded on the global growth of the service sector and its significance for society as a whole and for the individual human being. He has acted as session leader at many international conferences and is reviewer for several international conferences and journals. Dr. Elm has extensive experience in developing courses and teaching material for blended learning. He is currently co-supervising two doctoral students.